\documentclass{elsart}
\usepackage{epsfig}

\begin{document}

\begin{frontmatter}

\title{A characteristic frequency of two mutually interacting gas bubbles in an 
acoustic field}
\author{Masato Ida}
\address{Satellite Venture Business Laboratory, Gunma University, \\
1--5--1 Tenjin-cho, Kiryu-shi, Gunma 376--8515, Japan 
}

\begin{abstract}
Transition frequencies of two spherical gas bubbles interacting in an 
acoustic field are discussed theoretically. In the present study, 
transition frequency is defined as the frequency of 
external sound for which the phase difference between a bubble's pulsation 
and the external sound is $\pi / 2$. It is shown by a linear theory that a 
bubble interacting with a neighboring bubble has three (or fewer) 
transition frequencies but only two natural frequencies. This result means that 
the bubble has a characteristic frequency besides the natural frequencies.
\end{abstract}


\end{frontmatter}

It is known that radiative interaction between gas bubbles varies their 
natural (or resonance) frequencies, around which the bubbles indicate 
resonance response. This variation in the natural frequencies has been the 
subject of many studies 
\cite{ref1,ref2,ref3,ref4,ref5,ref6,ref7,ref8,ref9,ref10,ref11,ref12,ref13}. 
In a case where two spherical gas 
bubbles of different radii interact acoustically, a bubble has two (or one) 
natural frequencies (see, e.g., \cite{ref2,ref5,ref6,ref8}).

The aim of this Letter is to show that a gas bubble interacting with a 
neighboring bubble has an alternative characteristic frequency; we now call 
this a transition frequency. In the present work, a transition frequency is 
defined as the frequency of an 
external sound (the driving frequency) for which the phase difference 
between a bubble's pulsation and the external sound becomes $\pi / 2$. It is 
well known that, in a single-bubble case, the phase difference between the 
bubble and an external sound becomes $\pi / 2$ when the driving frequency is 
equal to the bubble's natural frequency \cite{ref14}. However, it is shown in 
this Letter that, in a double-bubble case, the number of 
transition frequencies differs from the number 
of natural frequencies. This result means that the bubbles in the latter 
case have a characteristic frequency that differs from the natural 
frequency.

First, we derive a mathematical model to use for the present investigation. 
This model describes the pulsation of two coupled bubbles. Since we are 
interested only in the qualitative natures of the transition frequencies, 
we use a classical 
theoretical model, called a coupled-oscillator or self-consistent model 
\cite{ref10,ref11,ref12}, although models with greater accuracy have been 
proposed \cite{ref4,ref6,ref7,ref9}.

A gas bubble immersed in a liquid pulsates when a sound wave is applied. The 
sound pressure at the bubble position drives the pulsation. When other 
bubbles (named ``bubble 2'' $\sim $ ``bubble $N$'', where $N$ is the total 
number of bubbles) exist near the bubble (bubble 1), the sound waves scattered 
by the other bubbles also drive the pulsation of bubble 1. Namely, the driving 
pressure acting on bubble 1, $p_{{\rm d}\,1}$, is expressed as
\begin{equation}
\label{eq1}
p_{{\rm d}\,1} = p_{{\rm ex}} + {\sum\limits_{j = 2}^{N} {p_{{\rm 
s}\,1j}} } {\rm ,}
\end{equation}
where $p_{\rm ex}$ and $p_{{\rm s}\,1j}$ are the sound pressures 
of the external sound field and the scattered wave emitted by bubble $j$, 
respectively, at the position of bubble 1. When the wavelength of the 
external sound wave is much larger than the bubbles' radii and the distances 
between the bubbles, $p_{{\rm ex}}$ can be considered spatially uniform. 
By assuming that the 
sphericity of the bubbles is thoroughly maintained and that the liquid 
surrounding the bubbles is incompressible, the scattered pressure can be 
estimated with (see, e.g., \cite{ref15})
\begin{equation}
\label{eq2}
p_{{\rm s}\,1j} \approx \frac{\rho }{{r_{1j}}}\frac{d}{{dt}}(R_j^2 \dot R_j),
\end{equation}
where $\rho$ is the density of the liquid, $r_{1j}$ is the distance between 
the centers of bubble 1 and bubble $j$, $R_j$ is the radius of 
bubble $j$, and the over dot denotes the time derivative.

The radial oscillation of a spherical bubble in an incompressible viscous 
liquid is described by the RPNNP equation \cite{ref16}:
\begin{equation}
\label{eq3}
R_1 \ddot R_1 + \frac{3}{2}\dot R_1^2 - \frac{1}{\rho }p_{w\,1} = - \frac{1}{\rho }p_{{\rm d}\,1} ,
\end{equation}
\[
p_{w\,1} = \left({P_0 + \frac{{2\sigma }}{{R_{10} }}} \right)\left( {\frac{{R_{10} }}{{R_1 }}} \right)^{3\kappa } - \frac{{2\sigma }}{{R_1 }} - \frac{{4\mu \dot R_1 }}{{R_1 }} - P_0 ,
\]
where $P_0 $ is the static pressure, $\sigma$ is the 
surface-tension coefficient at the bubble surface, $\mu $ is the viscosity of 
the liquid, $\kappa$ is the polytropic exponent of the gas inside the bubbles, 
and $R_{10}$ is the equilibrium radius of bubble 1. Substituting 
Eqs.~(\ref{eq1}) and (\ref{eq2}) into Eq.~(\ref{eq3}) yields
\[
R_1 \ddot R_1  + \frac{3}{2}\dot R_1^2  - \frac{1}{\rho }p_{w\,1}  =  - \frac{1}{\rho }[p_{{\rm ex}}  + \sum\limits_{j = 2}^N {\frac{\rho }{{r_{1j} }}\frac{d}{{dt}}(R_j^2 \dot R_j )} ] .
\]
When $N = 2$, this equation is reduced to
\begin{equation}
\label{eq4}
R_1 \ddot R_1 + \frac{3}{2}\dot R_1^2 - \frac{1}{\rho }p_{w\,1} = - \frac{1}{\rho }[p_{{\rm ex}} + \frac{\rho }{D}\frac{d}{{dt}}(R_2^2 \dot R_2 )],
\end{equation}
where $D = r_{12}$ ($ = r_{21}$). Exchanging subscripts 1 and 2 yields the 
model equation for bubble 2:
\begin{equation}
\label{eq5}
R_2 \ddot R_2 + \frac{3}{2}\dot R_2^2 - \frac{1}{\rho }p_{w\,2} = - \frac{1}{\rho }[p_{{\rm ex}} + \frac{\rho }{D}\frac{d}{{dt}}(R_1^2 \dot R_1 )] ,
\end{equation}
\[
p_{w\,2} = \left( {P_0 + \frac{{2\sigma }}{{R_{20} }}} \right)\left( {\frac{{R_{20} }}{{R_2 }}} \right)^{3\kappa }  - \frac{{2\sigma }}{{R_2 }} - \frac{{4\mu \dot R_2 }}{{R_2 }} - P_0.
\]
By assuming that the time-dependent bubble radii can be represented as 
$R_1 = R_{10} + e_1$, $R_2 = R_{20} + e_2$, and 
$\left| {e_1} \right| \ll R_{10}$, $\left| {e_2} \right| \ll R_{20}$, 
Eqs.~(\ref{eq4}) and (\ref{eq5}) are reduced to the following linear formulae:
\begin{equation}
\label{eq6}
\ddot e_1 + \omega _{10}^2 e_1 + \delta _1 \dot e_1 = - \frac{{p_{{\rm ex}} }}{{\rho R_{10} }} - \frac{{R_{20}^2 }}{{R_{10} D}}\ddot e_2,
\end{equation}
\begin{equation}
\label{eq7}
\ddot e_2 + \omega _{20}^2 e_2 + \delta _2 \dot e_2 = - \frac{{p_{{\rm ex}} }}{{\rho R_{20} }} - \frac{{R_{10}^2 }}{{R_{20} D}}\ddot e_1,
\end{equation}
where
\[
\omega _{j0}  = \sqrt {\frac{1}{{\rho R_{j0}^2 }}\left[ {3\kappa P_0  + (3\kappa  - 1)\frac{{2\sigma }}{{R_{j0} }}} \right]} 
\quad {\rm for} \;\; j = 1,2,
\]
are the partial natural (angular) frequencies of the bubbles, and
\[
\delta _j  = \frac{{4\mu }}{{\rho R_{j0}^2 }}
\quad {\rm for} \;\; j = 1,2,
\]
are the damping coefficients. (In general, the damping coefficients are 
determined by the sum of viscous, radiation, and thermal damping \cite{ref17}; 
we however use the above setting to simplify the following discussion.)

The system of equations (\ref{eq6}) and (\ref{eq7}) corresponds to that 
derived by Shima when $\sigma = 0$, $\delta _1 = 0$, and $\delta _2 = 0$ 
\cite{ref2}, and to that by Zabolotskaya when $\sigma = 0$ \cite{ref5}. 
This kind of system of differential equations, called the 
coupled-oscillator model or self-consistent model, has been used repeatedly 
to analyze acoustic properties of coupled bubbles 
\cite{ref2,ref5,ref8,ref10,ref11,ref12,ref13}, and has 
been proved to be a useful model that can provide a qualitatively accurate 
result despite its simple form.

Let us analyze in detail the transition frequencies of the linear coupled 
system. By assuming that 
the external sound pressure at the bubble position is written in the form of 
$p_{{\rm ex}}  =  - P_a \sin \omega t$, a harmonic 
steady-state solution of the linear coupled system is given as
\begin{equation}
\label{eq8}
e_1  = K_1 \sin (\omega t - \phi _1 ) ,
\end{equation}
where
\begin{equation}
\label{eq9}
K_1  = \frac{{P_a }}{{R_{10} \rho }}\sqrt {A_1^2  + B_1^2 } ,
\end{equation}
\[
\phi _1  = \tan ^{ - 1} \left( {\frac{{B_1 }}{{A_1 }}} \right) ,
\]
with
\begin{equation}
\label{eq10}
A_1 = \frac{{H_1 F + M_2 G}}{{F^2  + G^2 }} ,
\end{equation}
\begin{equation}
\label{eq11}
B_1 = \frac{{H_1 G - M_2 F}}{{F^2  + G^2 }} ,
\end{equation}
\[
F = L_1 L_2  - \frac{{R_{10} R_{20} }}{{D^2 }}\omega ^4  - M_1 M_2 ,
\]
\[
G = L_1 M_2  + L_2 M_1 ,
\quad
H_1  = L_2  + \frac{{R_{20} }}{D}\omega ^2 ,
\]
\[
L_1  = (\omega _{10}^2  - \omega ^2 ) ,
\quad
L_2  = (\omega _{20}^2  - \omega ^2 ) ,
\]
\[
M_1  = \delta _1 \omega ,
\quad
M_2  = \delta _2 \omega .
\]
Here the solution for only bubble 1 is shown. The phase difference 
$\phi _1$ becomes $\pi /2$ when
\begin{equation}
\label{eq12}
A_1  = 0 .
\end{equation}
It should be noted that a case in which both $A_1$ and $B_1$ become zero does 
not exist. From Eqs.~(\ref{eq10}) and (\ref{eq11}), one obtains
\[
A_1^2  + B_1^2  = \frac{{H_1 ^2  + M_2 ^2 }}{{F^2  + G^2 }}\left( {\frac{{P_a }}{{R_{10} \rho }}} \right)^2 .
\]
The numerator of this equation always has a nonzero value, since 
$M_2  > 0$; this result denies 
the existence of a case where both $A_1 = 0$ and $B_1 = 0$ are true. Also, 
it should be noted that $F^2 + G^2$ appearing in the denominator of 
Eq.~(\ref{eq10}) always has a nonzero value. When $G = 0$, $F$ is reduced to
\[
F = - \frac{{M_2 }}{{M_1 }}L_1^2 - \frac{{R_{10} R_{20} }}{{D^2 }}\omega ^4 - M_1 M_2 .
\]
This has a nonzero, negative value because $M_2 L_1^2 /M_1 \ge 0$, 
$R_{10} R_{20} \omega ^4 /D^2 > 0$, and $M_1 M_2 > 0$. This result means that 
no case exists where both $F = 0$ and $G = 0$ are true. As a consequence, 
Eq.~(\ref{eq12}) is reduced to
\begin{equation}
\label{eq13}
H_1 F + M_2 G = 0 .
\end{equation}
In the following, we analyze this equation.

When the damping terms in Eq.~(\ref{eq13}) are negligible (but exist), one can 
easily obtain the solution for this equation. By assuming that 
$M_1 \approx 0$ and $M_2 \approx 0$, one obtains
\[
H_1 F \approx 0 .
\]
This equation can be rewritten into two independent equations:
\begin{equation}
\label{eq14}
F \approx L_1 L_2  - \frac{{R_{10} R_{20} }}{{D^2 }}\omega ^4 = 0
\end{equation}
and
\begin{equation}
\label{eq15}
H_1 = L_2  + \frac{{R_{20} }}{D}\omega ^2 = 0 .
\end{equation}
Equation (\ref{eq14}) is the same as the theoretical formula given in 
Refs.~\cite{ref2,ref5} to derive the natural frequencies of a double-bubble 
system; 
namely, the transition frequencies given by this equation correspond to the 
natural frequencies. This equation predicts the existence of two natural 
frequencies,
\begin{equation}
\label{eq16}
\omega _{1 \pm }^2  = \frac{{\omega _{10}^2  + \omega _{20}^2  \pm \sqrt {\left( {\omega _{10}^2  - \omega _{20}^2 } \right)^2  + 4\omega _{10}^2 \omega _{20}^2 \frac{{R_{10} R_{20} }}{{D^2 }}} }}{{2\left( {1 - \frac{{R_{10} R_{20} }}{{D^2 }}} \right)}} ,
\end{equation}
and is symmetric; namely, it exchanges subscripts 1 and 2 (or 10 and 20) in 
this equation to reproduce the same equation. This means that the two 
bubbles have the same two natural frequencies. One of the natural 
frequencies converges to $\omega _{10}$ and the other to $\omega _{20}$ for 
$D \to \infty$, and the higher natural frequency rises and the lower one falls 
as $D$ decreases \cite{ref2,ref18}.

The solution of Eq.~(\ref{eq15}) given for the first time is
\begin{equation}
\label{eq17}
\omega _1^2  = \frac{{\omega _{20}^2 }}{{1 - R_{20} /D}} .
\end{equation}
This converges to $\omega _{20}^2$ for $D \to \infty$, and increases 
as $D$ decreases. In contrast to Eq.~(\ref{eq14}), Eq.~(\ref{eq15}) is 
asymmetric ($H_1 \ne H_2$); namely, this serves to break the symmetry of 
the natural frequency mentioned above. The transition frequency given by 
Eq.~(\ref{eq17}) is not the 
natural frequency because it cannot be given by the natural frequency 
analysis performed in Refs.~\cite{ref2,ref5}. (Even the other models used in, 
e.g., \cite{ref6,ref8} give only two natural frequencies.)

The results given above show that, when the damping effect is negligible and 
the radii of bubbles are not equal, the bubbles have three asymmetric 
transition frequencies; one of these, for $D \to \infty$, converges to the 
partial natural frequency of a corresponding 
bubble, while the remaining two converge to that of a neighboring bubble. 
(In the following, the former transition frequency is called 
``fundamental transition frequency'' (FTF), and the latter two are called 
``sub transition frequencies'' (STFs).) One of the STFs always increases as 
bubbles approach each 
other. The other STF decreases (increases) and the FTF increases (decreases) 
when the partial natural frequency of the bubble is higher (lower) than that 
of the neighboring bubble.

\begin{figure}
\begin{center}
\epsfxsize = 12 cm
\epsffile{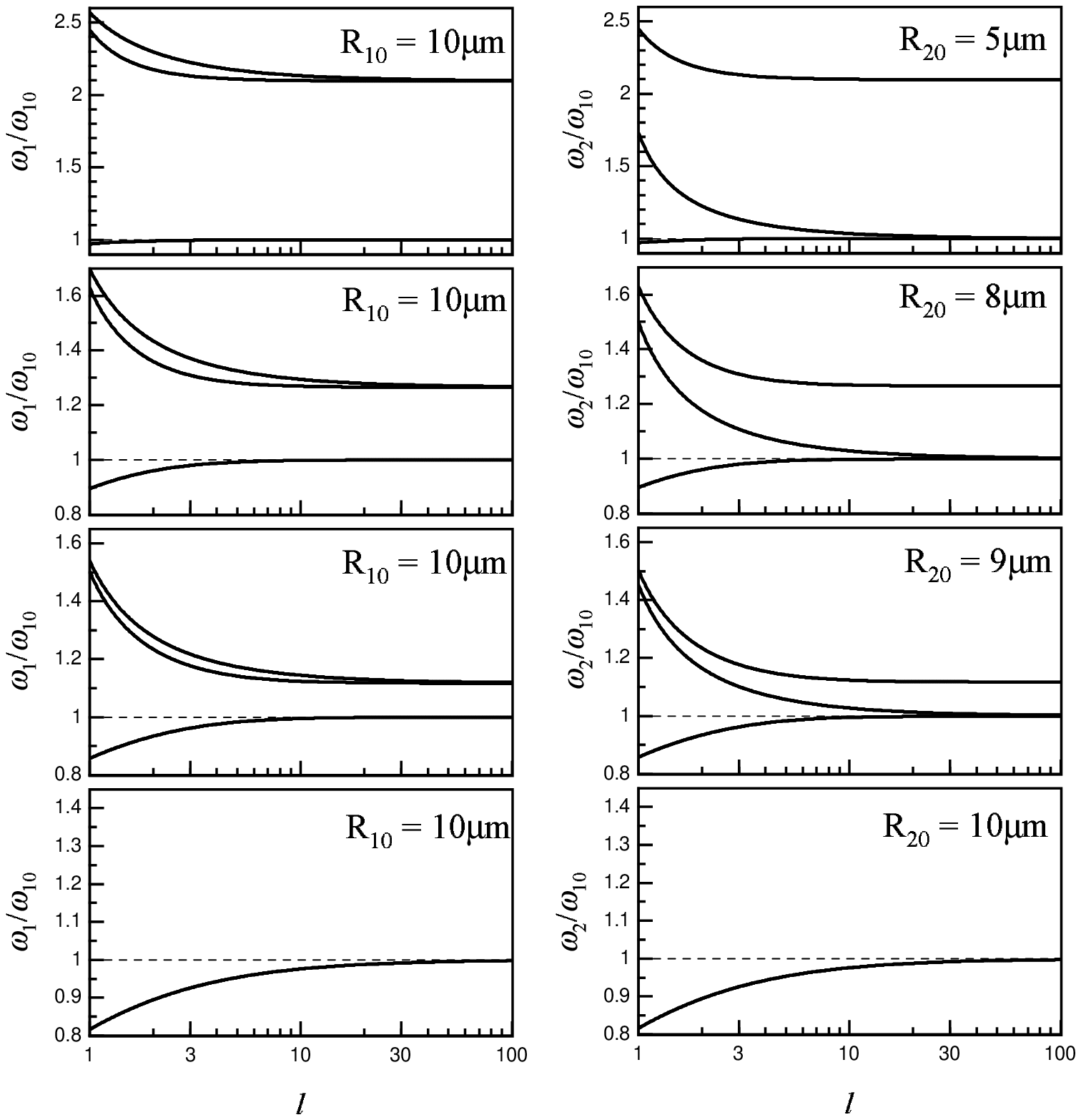}
\caption{
Transition frequencies of bubbles 1 ($\omega _1$) and 2 ($\omega _2$) for 
$R_{j0} \sim 10$ $\mu$m and $\mu \approx 0$, normalized by $\omega _{10}$. 
The dashed line denotes $\omega _j /\omega _{10} = 1$. The highest 
transition frequency of bubble 1 and the second highest one of bubble 2 are 
given by Eq.~(\ref{eq17}).}
\label{fig1}
\end{center}
\end{figure}

Figure \ref{fig1} shows the transition frequencies for $\mu \approx 0$ as a 
function of $l = D/(R_{10} + R_{20})$. Other parameters are set to 
$\rho = 1000$ kg/m$^3$, $P_0 = 1$ atm, $\sigma = 0.0728$ N/m, and 
$\kappa = 1.4$. Equations (\ref{eq16}) and (\ref{eq17}) are used in plotting 
those graphs. The radius of bubble 1 is fixed to 
10 $\mu$m and that of bubble 2 varies from 5 $\mu$m to 10 $\mu$m. As 
discussed above, three transition frequencies changing with $D$ appear in 
each graph, except for the case of $R_{10} = R_{20}$, where only one 
decreasing transition frequency appears. (When the two bubbles have the same 
radii and pulsate in phase, Eq.~(\ref{eq6}) is reduced to
\[
\left( {1 + \frac{{R_{10} }}{D}} \right)\ddot e_1  + \omega _{10}^2 e_1 + \delta _1 \dot e_1  =  - \frac{{p_{{\rm ex}} }}{{\rho R_{10} }} .
\]
This equation has only one transition frequency \cite{ref5,ref10} of
\begin{equation}
\label{eq18}
\omega _1^2  = \frac{{\omega _{10}^2 }}{{1 + R_{10} /D}} ,
\end{equation}
which converges to $\omega _{10}^2$ for $D \to \infty$ and decreases 
with decreasing $D$.) It is interesting to point out that the larger bubble, 
whose partial natural 
frequency is lower than that of the smaller bubble, has the highest 
transition frequency among all the bubbles.

\begin{figure}
\begin{center}
\epsfxsize = 12 cm
\epsffile{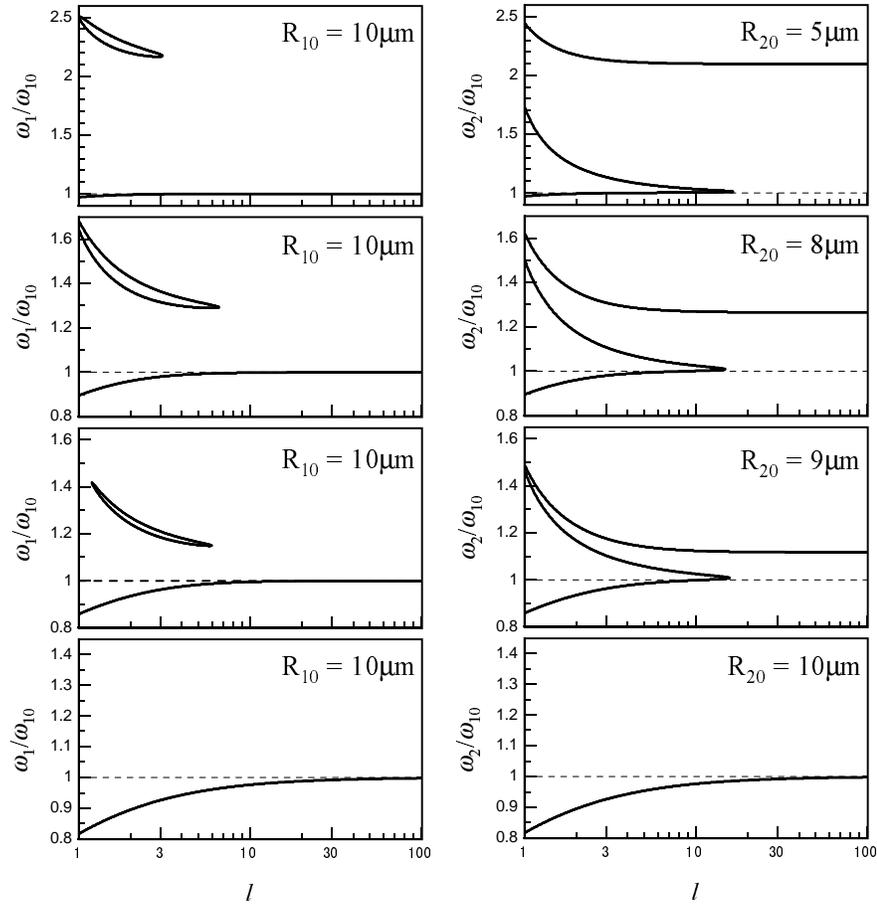}
\caption{
Transition frequencies for $R_{j0} \sim 10$ $\mu$m normalized by 
$\omega _{10}$. The viscous effect is taken into account. The dashed line 
denotes $\omega _j /\omega _{10} = 1$.}
\label{fig2}
\end{center}
\end{figure}

\begin{figure}
\begin{center}
\epsfxsize = 12 cm
\epsffile{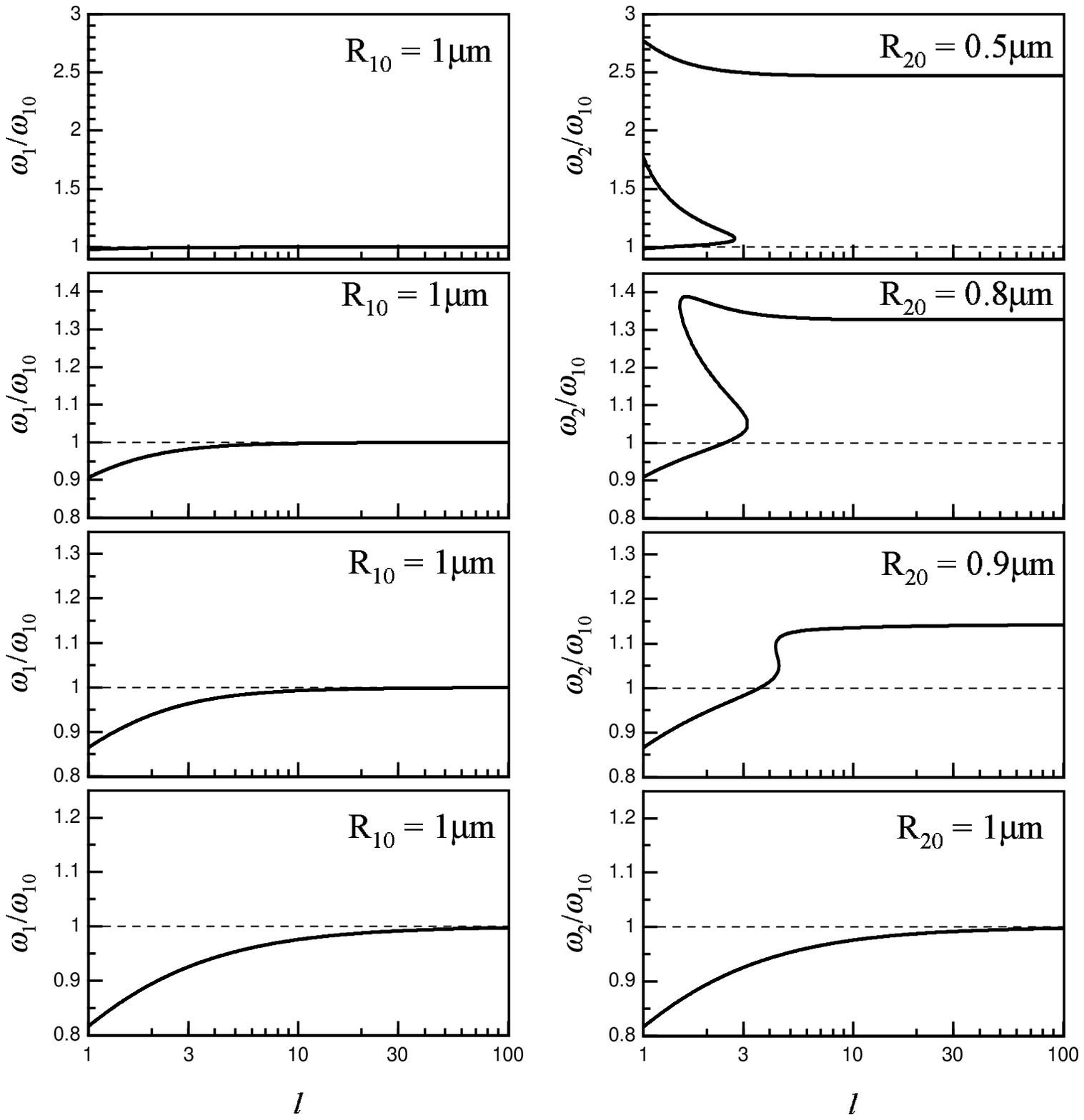}
\caption{
Same as Fig.~\ref{fig2} except for $R_{j0} \sim 1$ $\mu$m.}
\label{fig3}
\end{center}
\end{figure}

\begin{figure}
\begin{center}
\epsfxsize = 12 cm
\epsffile{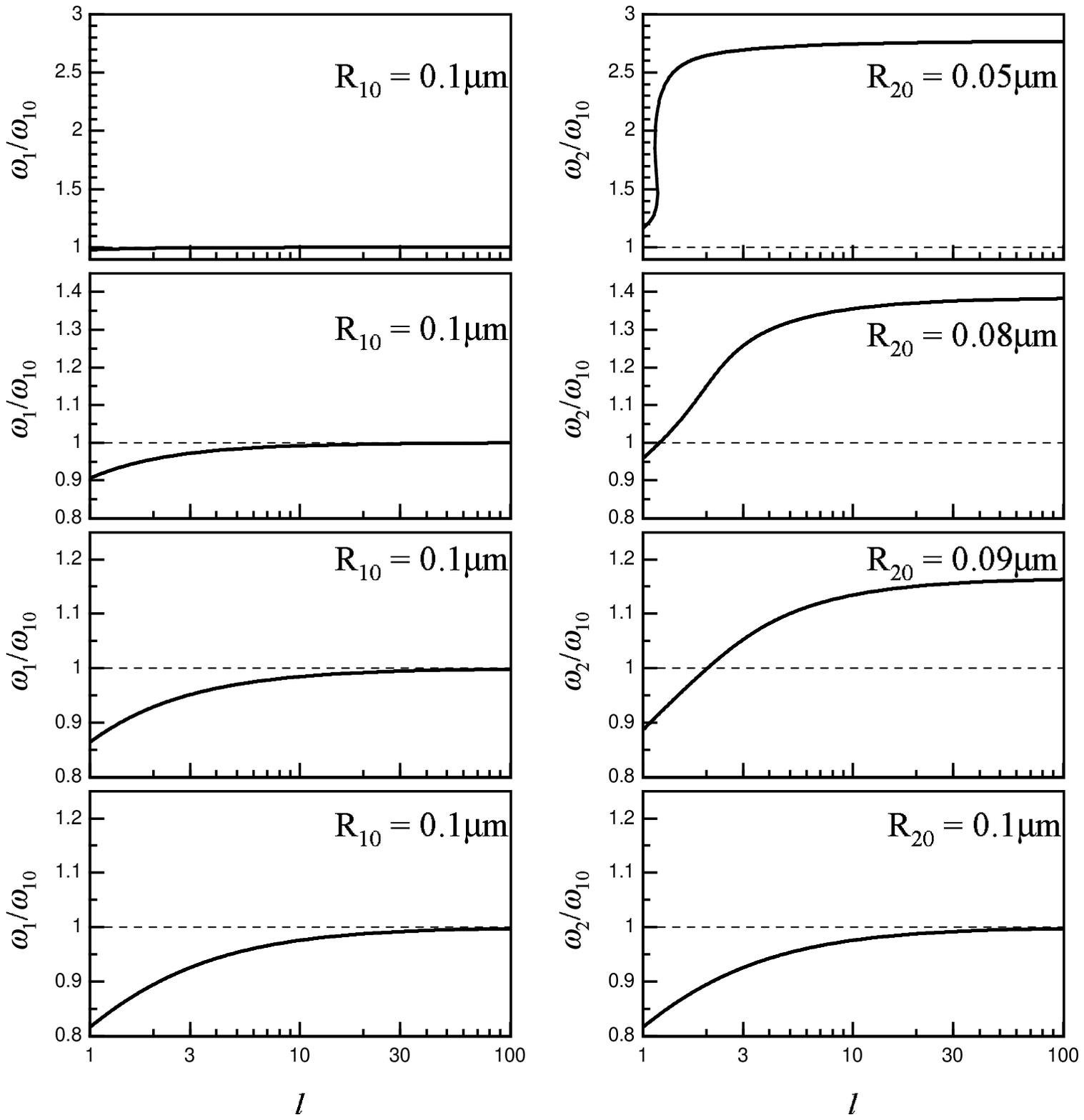}
\caption{
Same as Fig.~\ref{fig2} except for $R_{j0} \sim 0.1$ $\mu$m.}
\label{fig4}
\end{center}
\end{figure}

Now we present numerical solutions of Eq.~(\ref{eq13}) to examine the 
influences of viscosity on transition frequencies. Figures 
\ref{fig2}$-$\ref{fig4} show the results obtained by using 
$\mu = 1.137 \times 10^{-3}$ kg/(m s), which corresponds to the viscosity of 
water at room temperature. From those figures, we can observe that, as the 
viscous effect grows strong, i.e., the bubbles' radii become smaller, the 
STFs vanish gradually from the large-distance region (and sometimes from the 
small-distance region), and only the FTF remains. In the case of 
$R_{10} = 1$ $\mu$m, the STFs of a larger bubble disappear, and, when 
the bubbles are of similar size to each other, the FTF and the higher STF of a 
smaller bubble vanish in the small-distance region. In the case of 
$R_{10} = 0.1$ $\mu$m, it is difficult to distinguish the STFs from the FTF 
since only a smooth curve, decreasing almost monotonically with decreasing $l$, 
appears. (The transition frequency of the smaller bubble remaining in 
the small-distance region may be the lower one of the STFs.) Those results 
show that the influence of the mutual interaction of the bubbles on the 
transition frequencies depends on the viscosity of the surrounding material, 
and that this 
interaction weakens as the viscous effect strengthens, i.e., the threshold 
of the distance for the appearance of the STFs is shortened.

\begin{figure}
\begin{center}
\epsfxsize = 8 cm
\epsffile{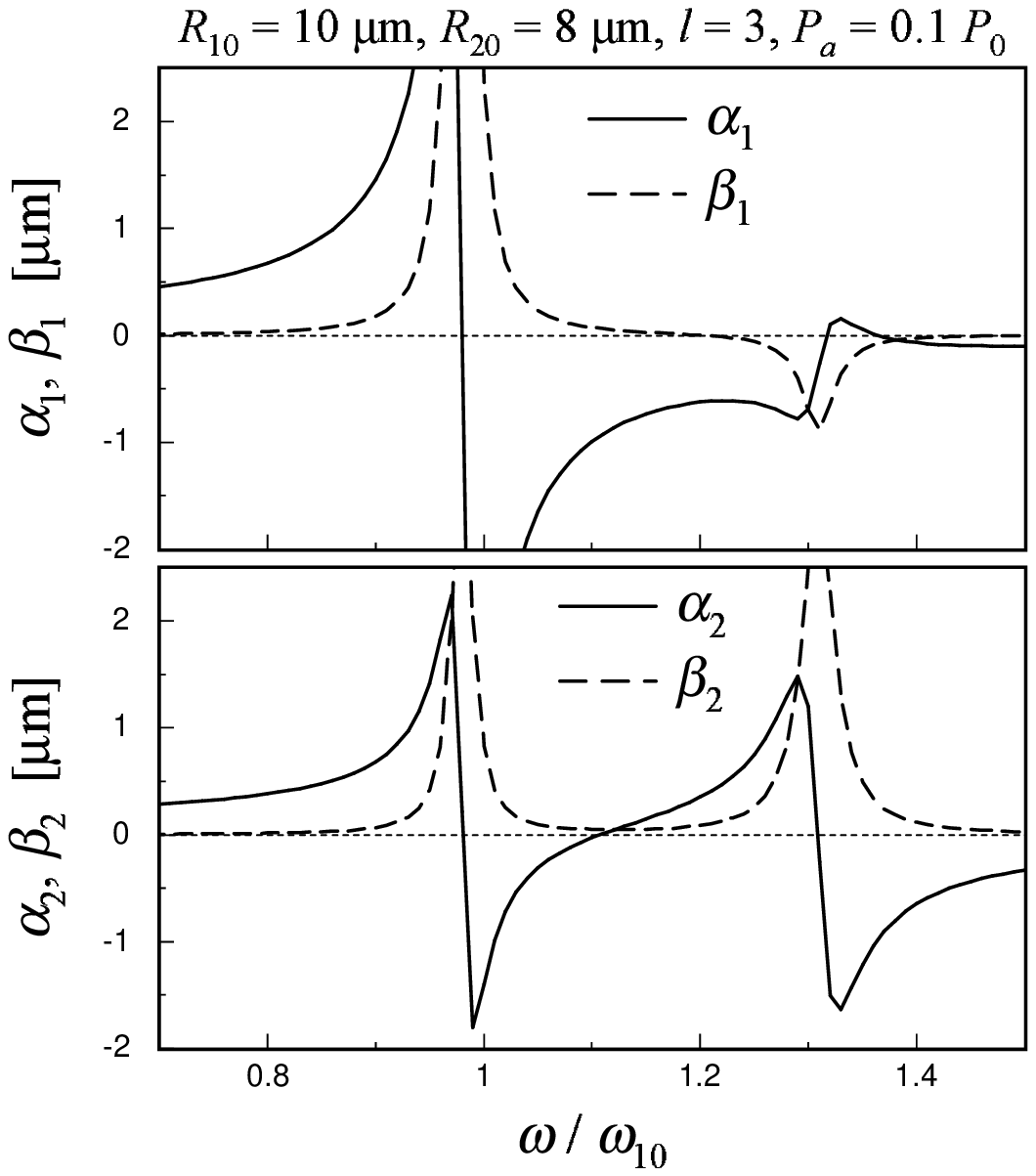}
\caption{
$\alpha _j  = (P_a /R_{j0} \rho )A_j$ and 
$\beta _j = (P_a /R_{j0} \rho )B_j$ for $R_1 = 10$ $\mu$m, $R_2 = 8$ $\mu$m, 
and $l = 3$.}
\label{fig5}
\end{center}
\end{figure}

Lastly, we briefly discuss the pulsation amplitudes of the bubbles. Figure 
\ref{fig5} shows $\alpha _j  \equiv (P_a /R_{j0} \rho )A_j$ and 
$\beta _j  \equiv (P_a /R_{j0} \rho )B_j$ for $R_1 = 10$ $\mu$m, 
$R_2 = 8$ $\mu$m, $l = 3$, and $\mu = 1.137 \times 10^{-3}$ kg/(m s) as 
functions of $\omega /\omega _{10}$, where we set $P_a = 0.1 P_0$. The present 
setting is equivalent to that used for one of the graphs shown in 
Fig.~\ref{fig2} except for $l$. In each graph shown in Fig.~\ref{fig5}, we 
observe three reversals of the sign of $\alpha _j$ and only two 
resonance responses, as expected. At the transition frequency that does not 
correspond to the natural frequency (the highest one of bubble 1 and the 
second highest one of bubble 2), no resonance response is obtained. This 
result confirms that the characteristic frequency given by Eq.~(\ref{eq15}) is 
not the natural frequency.

In summary, it was predicted in this Letter that a gas bubble interacting 
acoustically with a neighboring bubble has three transition frequencies that 
make the phase difference between the bubble's pulsation and an external sound 
be $\pi /2$, while readymade 
theories predict only two natural frequencies. This present result shows 
that, in a double-bubble case, the phase reversal of a bubble (e.g., from 
in-phase to out-of-phase with an external sound) can take place not only at 
the bubble's natural frequencies but also at some other frequency. The 
present theory for transition frequencies may be useful for understanding the 
reversal of the sign of the secondary Bjerknes force, which has so far been 
interpreted using only the natural frequencies \cite{ref5,ref19,ref20,add01}. 
(This subject is discussed in a separate paper of ours \cite{ref21}.) 
Furthermore, the present results might also 
affect understandings of some related subjects such as acoustic localization 
\cite{ref22,ref23} and superresonances \cite{ref10}, previous investigations of 
which have employed systems containing only identical bubbles. The present 
theory for a double-bubble system will be verified \cite{ref24} by the 
direct numerical simulation technique proposed in Ref.~\cite{ref25}, and will 
be extended to a theory for an $N$-bubble system \cite{ref26}, where $N$ 
denotes an arbitrary positive integer.

\newpage

\newpage

\newpage

\newpage

\newpage

\begin{thebibliography}{}
\bibitem{ref1}
M. Strasberg, J. Acoust. Soc. Am. \textbf{25}, 536 (1953).
\bibitem{ref2}
A. Shima, Trans. ASME, J. Basic Eng. \textbf{93}, 426 (1971).
\bibitem{ref3}
G. N. Kuznetsov and I. E. Shchekin, Sov. Phys. Acoust. \textbf{21}, 147 
(1975).
\bibitem{ref4}
J. F. Scott, J. Fluid Mech. \textbf{113}, 487 (1981).
\bibitem{ref5}
E. A. Zabolotskaya, Sov. Phys. Acoust. \textbf{30}, 365 (1984).
\bibitem{ref6}
Yu. A. Kobelev and L. A. Ostrovskii, Sov. Phys. Acoust. \textbf{30}, 427 
(1984).
\bibitem{ref7}
S. T. Zavtrak, Sov. Phys. Acoust. \textbf{33}, 145 (1987).
\bibitem{ref8}
H. Takahira, S. Fujikawa and T. Akamatsu, JSME Int. J. Ser. II 
\textbf{32} (1989) 163.
\bibitem{ref9}
A. S. Sangani, J. Fluid Mech. \textbf{232}, 221 (1991).
\bibitem{ref10}
C. Feuillade, J. Acoust. Soc. Am. \textbf{98}, 1178 (1995).
\bibitem{ref11}
Z. Ye and C. Feuillade, J. Acoust. Soc. Am. \textbf{102}, 798 (1997).
\bibitem{ref12}
C. Feuillade, J. Acoust. Soc. Am. \textbf{109}, 2606 (2001).
\bibitem{ref13}
P.-Y. Hsiao, M. Devaud, and J.-C. Bacri, Eur. Phys. J. E \textbf{4}, 5 
(2001).
\bibitem{ref14}
T. G. Leighton, \textit{The Acoustic Bubble} (Academic Press, London, 1994), p.293.
\bibitem{ref15}
R. Mettin, I. Akhatov, U. Parlitz, C. D. Ohl, and W. Lauterborn, Phys. 
Rev. E \textbf{56}, 2924 (1997).
\bibitem{ref16}
W. Lauterborn, J. Acoust. Soc. Am. \textbf{59}, 283 (1976).
\bibitem{ref17}
C. Devin, J. Acoust. Soc. Am. \textbf{31}, 1654 (1959).
\bibitem{ref18}
M. Ida, e-Print, physics/0108067 (not for submission).
\bibitem{ref19}
A. A. Doinikov and S. T. Zavtrak, Phys. Fluids \textbf{7}, 1923 (1995).
\bibitem{ref20}
A. A. Doinikov and S. T. Zavtrak, J. Acoust. Soc. Am. \textbf{99}, 3849 
(1996).
\bibitem{add01}
A. Harkin, T. J. Kaper, and A. Nadim, J. Fluid Mech. \textbf{445}, (2001) 377.
\bibitem{ref21}
M. Ida, (submitted); e-Print, physics/0109005.
\bibitem{ref22}
Z. Ye and A. Alvarez, Phys. Rev. Lett. \textbf{80}, 3503 (1998).
\bibitem{ref23}
A. Alvarez and Z. Ye, Phys. Lett. A \textbf{252}, 53 (1999).
\bibitem{ref24}
M. Ida, e-Print, physics/0111138.
\bibitem{ref25}
M. Ida and Y. Yamakoshi, Jpn. J. Appl. Phys. \textbf{40}, 3846 (2001).
\bibitem{ref26}
M. Ida, (submitted); e-Print, physics/0108056.
\end{thebibliography}
\end{document}